# Projectile motion in a medium with quadratic drag at constant horizontal wind

Peter Chudinov

*Department of Engineering, Perm State Agro-Technological University, 614990, Perm, Russia*

E-mail: chupet@mail.ru

A classic problem of the motion of a projectile thrown at an angle to the horizon is studied. Air resistance force is taken into account with the use of the quadratic resistance law. The action of the wind is also taken into account, which is considered constant and horizontal. The projectile velocity hodograph equation is used to account for the effect of wind. Comparatively simple analytical approximations are proposed for the main variables of motion (cartesian projectile coordinates and time). All obtained formulas contain only elementary functions. The proposed formulas are universal, that is, they can be used for any initial conditions of throwing. In addition, they have acceptable accuracy over a wide range of the change of parameters. The motion of a golf ball, a tennis ball and shuttlecock of badminton are presented as examples. The calculation results show good agreement between the proposed analytical solutions and numerical solutions. The proposed analytical formulas can be useful for all researchers of this classical problem.

## 1. Introduction

The study of the motion of a projectile, thrown at an angle to the horizon, is a wonderful classical problem. This issue has been the subject of great interest for investigators for centuries. Currently, the study of parabolic motion, in the absence of any drag force, is a common example in introductory physics courses at all universities. The theory of parabolic motion allows you to analytically determine the trajectory and all important characteristics of the movement of the projectile. Introduction of air resistance forces into the study of the motion, however, complicates the problem and makes it difficult to obtain analytical solutions. This especially applies to the movement of the projectile, subjected to quadratic air drag force. The problem becomes even more complicated if the effect of wind is taken into account when the projectile moves in a medium with quadratic resistance. For a quadratic or more general non-linear drag force, results have been obtained only numerically. The detailed analysis of wind-influenced projectile motion in the case of linear and nonlinear (quadratic or nonquadratic) drag force is reviewed in [6]. From an educational point of view, many students are not confident with numerical methods and gain little insight from using them. So the description of the projectile motion by means of simple approximate analytical formulas under air resistance and wind has great methodological and educational importance.

    Many researches are devoted to the numerical study of projectile motion in the presence of quadratic resistance of the medium and the action of the wind, for example [1 – 3, 7 – 12]. This is due to the fact that the differential equations of motion do not allow an analytical solution, although in the case of quadratic resistance there is a closed relationship between the velocity and the corresponding angular parameter (the velocity hodograph equation). Only for the linear resistance force, closed-form solutions have been obtained [2], including expressions for the change in time of the velocity components and the shape of the trajectory. In this case, the

special Lambert function $W$ is used. Thus, the problem of the analytical description of the projectile motion with quadratic resistance and taking into account the wind remains relevant. In this paper, an attempt is made to obtain analytical approximations for solving this problem, at least approximate ones. The main goal of this work is to give analytical approximations for the projectile trajectory as simple as possible from a technical point of view, in order to be grasped even by first-year undergraduates. The proposed formulas contain only elementary functions, which makes it easier for students to understand the problem.

The research is based on two previously solved problems. The first task is to obtain a hodograph of the projectile velocity with a quadratic resistance of the medium, but without taking into account the wind. This problem has been solved and described long ago [11]. Accounting for the action of a constant wind in the velocity hodograph equation was made in the paper [10]. The velocity hodograph equation taking into account the wind is used in this study. The second task is to successfully approximate the complex transcendental function that enters the hodograph equation and the equations of motion, and apply this approximation to solve the equations of motion of the projectile. Without taking into account the wind, this was done in [4]. The idea of [4] is used in this study, taking into account the adopted wind model. The conditions of applicability of the quadratic resistance law are deemed to be fulfilled, i.e. Reynolds number $Re$ lies within $1 \times 10^3 < Re < 2 \times 10^5$. Magnus forces are not included in this work.

## 2. Equations of projectile motion and velocity hodograph equation

Here we state the formulation of the problem and the equations of the motion [4]. Let us consider the motion of a projectile with mass $m$ launched at an angle $\theta_0$ with an initial speed $V_0$ under the influence of the force of gravity and resistance force $R = mgkV^2$. Here $g$ is the acceleration of gravity, $k$ is the drag constant and $V$ is the speed of the object. Air resistance force $R$ is proportional to the square of the speed of the projectile and is directed opposite the velocity vector (see Fig. 1). The impact of the wind on the projectile will be modeled as a constant horizontal speed of the air $\overline{w}$. It is assumed that the projectile is at the origin at the initial instant and the point of impact of the projectile lies on the same horizontal $y = 0$ (see Fig. 1). In ballistics in the absence of wind, the movement of a projectile is often studied in projections on natural axes. The equations of the projectile motion in this case have the form

$$\frac{dV}{dt} = -g\sin\theta - gkV^2, \quad \frac{d\theta}{dt} = -\frac{g\cos\theta}{V}, \quad \frac{dx}{dt} = V\cos\theta, \quad \frac{dy}{dt} = V\sin\theta. \qquad (1)$$

Here $\theta$ is the angle between the tangent to the trajectory of the projectile and the horizontal, $x, y$ are the Cartesian coordinates of the projectile. The well-known solution [11] of system (1) consists of an explicit analytical dependence of the velocity on the slope angle of the trajectory (hodograph equation for projectile velocity) and three quadratures

$$V(\theta) = \frac{V_0 \cos\theta_0}{\cos\theta\sqrt{1 + kV_0^2 \cos^2\theta_0 \left(f(\theta_0) - f(\theta)\right)}}, \quad f(\theta) = \frac{\sin\theta}{\cos^2\theta} + \ln\tan\left(\frac{\theta}{2} + \frac{\pi}{4}\right), \qquad (2)$$

$$x = x_0 - \frac{1}{g}\int_{\theta_0}^{\theta} V^2 d\theta, \quad y = y_0 - \frac{1}{g}\int_{\theta_0}^{\theta} V^2 \tan\theta d\theta, \quad t = t_0 - \frac{1}{g}\int_{\theta_0}^{\theta} \frac{V}{\cos\theta} d\theta. \qquad (3)$$

Here $t_0$ is the initial value of the time, $x_0, y_0$ are the initial values of the coordinates of the projectile. In the following formulas, we take $x_0 = 0$, $t_0 = 0$, $y_0 \geq 0$. The drag coefficient $k$, used in formulas (1) – (2), can be calculated through the terminal speed: $k = 1/V_{term}^2$ [5].

Terminal velocity is the maximum velocity attainable by an object as it falls through a fluid (air is the most common example).

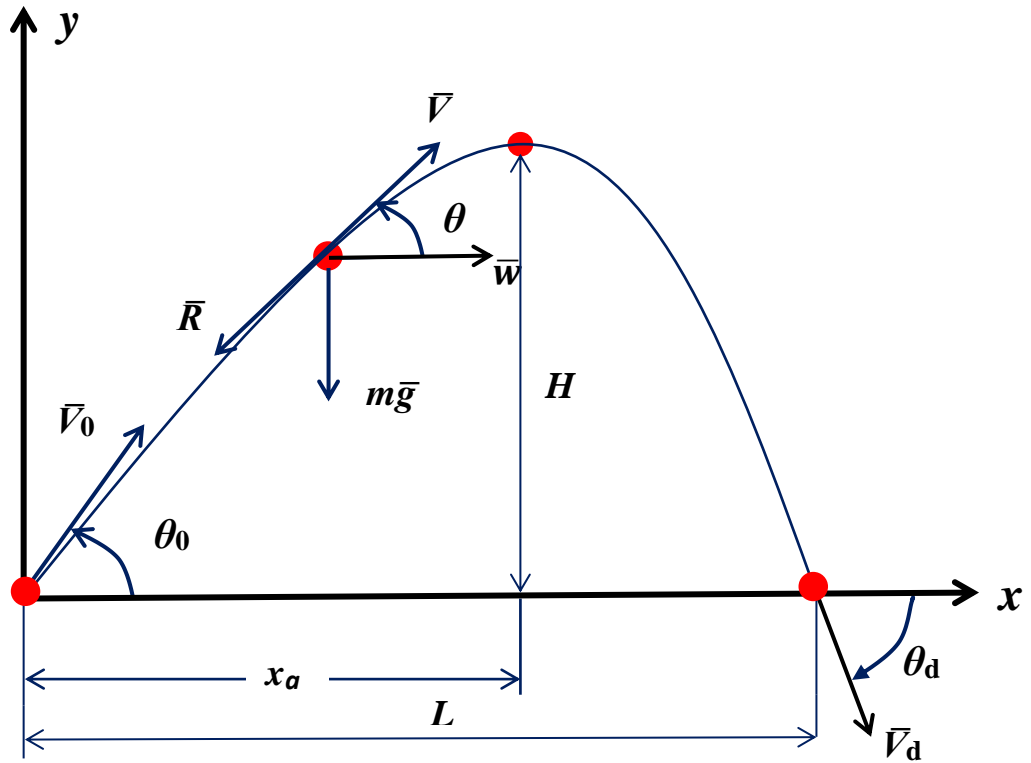

**Fig. 1.** Acting forces on the projectile and basic motion parameters.

Let us write the hodograph equation (3) and integrals (4) under the action of a constant horizontal wind, following review [6]. In the presence of horizontal wind whose velocity vector $\bar{w}$ has the components $(w, 0)$, the drag force in the model of quadratic resistance is

$$\bar{R} = -c|\bar{V} - \bar{w}|(\bar{V} - \bar{w}), \qquad (4)$$

where $c = gk$ is the drag coefficient. Then the differential equations of projectile motion in projections onto the Cartesian axes have the form

$$\frac{dx}{dt} = V_x, \quad \frac{dV_x}{dt} = -c\sqrt{(V_x - w)^2 + V_y^2}\,(V_x - w),$$

$$\frac{dy}{dt} = V_y, \quad \frac{dV_y}{dt} = -c\sqrt{(V_x - w)^2 + V_y^2}\,V_y - g. \qquad (5)$$

Initial conditions in this case $x_0 = y_0 = 0$, $V_x(0) = V_0 \cos\theta_0$, $V_y(0) = V_0 \sin\theta_0$. We introduce the relative velocity of the projectile by the equality $\bar{u} = \bar{V} - \bar{w}$ with components $u_x = V_x - w$, $u_y = V_y$. The magnitude of the relative velocity is $u = \sqrt{u_x^2 + u_y^2}$. Then equations (5) take the form

$$\frac{dx}{dt}=u_x+w, \quad \frac{du_x}{dt}=-c\sqrt{u_x^2+u_y^2}u_x, \quad \frac{dy}{dt}=u_y, \quad \frac{du_y}{dt}=-c\sqrt{u_x^2+u_y^2}u_y-g. \tag{6}$$

Equations (6) have the same form as when the projectile moves without wind [6]. Therefore, for equations (6), the relative velocity hodograph equation takes place, which is similar to equation (2):

$$u(\varphi)=\frac{u_0\cos\varphi_0}{\cos\varphi\sqrt{1+ku_0^2\cos^2\varphi_0\left(f(\varphi_0)-f(\varphi)\right)}}, \quad f(\varphi)=\frac{\sin\varphi}{\cos^2\varphi}+\ln\tan\left(\frac{\varphi}{2}+\frac{\pi}{4}\right). \tag{7}$$

The auxiliary angular parameter $\varphi$ is the angle between the relative velocity vector $\bar{u}$ and the positive axis $x$ and is determined by the formula [10]

$$\varphi=\arctan\left(\frac{u_y}{u_x}\right)=\arctan\left(\frac{V\sin\theta}{V\cos\theta-w}\right), \quad \varphi_0=\arctan\left(\frac{V_0\sin\theta_0}{V_0\cos\theta_0-w}\right),$$

$$u_0=\sqrt{u_x^2(0)+u_y^2(0)}=\sqrt{(V_0\cos\theta_0-w)^2+(V_0\sin\theta_0)^2}=\sqrt{V_0^2-2V_0w\cos\theta_0+w^2}.$$

Quadratures (3) in this case take the form

$$x=x_0-\frac{1}{g}\int_{\varphi_0}^{\varphi}u^2d\varphi-\frac{w}{g}\int_{\varphi_0}^{\varphi}\frac{u}{\cos\varphi}d\varphi, \quad y=y_0-\frac{1}{g}\int_{\varphi_0}^{\varphi}u^2\tan\varphi d\varphi, \quad t=t_0-\frac{1}{g}\int_{\varphi_0}^{\varphi}\frac{u}{\cos\varphi}d\varphi. \tag{8}$$

Since the dependence $u(\varphi)$ is known, the problem of projectile motion in a medium with quadratic resistance at a constant horizontal wind is reduced to the calculation of integrals (8).

### 3. Computational formulas of the problem

The main problem in calculating integrals (8) is that the complex form of the transcendental function $f(\varphi)$ in the velocity hodograph equation does not allow finding a solution in elementary functions. In this paper, we propose convenient solution of the problem taking into account the wind, similar to the solution in [4]. All mathematical calculations are omitted. According to the approach used in [4], we divide the entire range of the trajectory angle $\varphi_0\geq\varphi>-\pi/2$ into three intervals: $\varphi_0\geq\varphi\geq 0$, $0\geq\varphi\geq\varphi_1$, $\varphi_1\geq\varphi>-\pi/2$. The value $\varphi_1$ is determined by the equality $\varphi_1=-\frac{\varphi_0}{2}-\frac{\pi}{4}$. Such a partition allows one to construct a solution over the entire interval of angle change $\varphi$. The first interval corresponds to the projectile lifting stage, the other two intervals correspond to the descent stage.

An approximation of the function $f(\varphi)$ of the following form was proposed in work of [4]:

$$f_a(\varphi)=\alpha_1\tan\varphi+\alpha_2\tan^2\varphi, \quad \text{on condition} \quad \varphi\geq 0,$$

$$f_a(\varphi)=\alpha_1\tan\varphi-\alpha_2\tan^2\varphi, \quad \text{on condition} \quad \varphi\leq 0.$$

The coefficients $\alpha_1, \alpha_2$ are chosen in such a way that the functions $f(\varphi)$ and $f_a(\varphi)$ are closely related to each other. For this, the following conditions are used

$$f_a(\varphi_0) = f(\varphi_0), \qquad f'_a(\varphi_0) = f'(\varphi_0). \tag{9}$$

From conditions (9) we have

$$\alpha_1 = 2\cot\varphi_0 \ln\tan\left(\frac{\varphi_0}{2} + \frac{\pi}{4}\right), \qquad \alpha_2 = \frac{1}{\sin\varphi_0} - \frac{\alpha_1}{2}\cot\varphi_0.$$

Now integrals (8) can be calculated in elementary functions. Omitting intermediate calculations, we write down the final formulas.

On the first interval, the functions $x(\varphi), y(\varphi), t(\varphi)$ have the following form:

$$x_1(\varphi) = x_0 + \frac{2}{gk\alpha_1\Delta_1}\arctan\left(\frac{1+2b_2\tan\varphi}{\Delta_1}\right)\bigg|_{\varphi_0}^{\varphi} - \frac{w}{g\sqrt{k\alpha_2}}\arcsin\left(\frac{1+2b_2\tan\varphi}{\Delta_3}\right)\bigg|_{\varphi_0}^{\varphi},$$

$$y_1(\varphi) = y_0 - \frac{1}{gk\alpha_1\Delta_1 b_2}\arctan\left(\frac{1+2b_2\tan\varphi}{\Delta_1}\right)\bigg|_{\varphi_0}^{\varphi} + \frac{1}{2gk\alpha_1 b_2}\ln\left(b_2\tan^2\varphi + \tan\varphi - b_1\right)\bigg|_{\varphi_0}^{\varphi},$$

$$t_1(\varphi) = t_0 - \frac{1}{g\sqrt{k\alpha_2}}\arcsin\left(\frac{2b_2\tan\varphi+1}{\Delta_3}\right)\bigg|_{\varphi_0}^{\varphi}.$$

On the second interval we have

$$x_2(\varphi) = x_1(0) - x_1(\varphi_0) + \frac{2}{gk\alpha_1\Delta_2}\arctan\left(\frac{1-2b_2\tan\varphi}{\Delta_2}\right)\bigg|_0^{\varphi} - \frac{w}{g\sqrt{-k\alpha_2}}\arcsin\left(\frac{1-2b_2\tan\varphi}{\Delta_4}\right)\bigg|_0^{\varphi},$$

$$y_2(\varphi) = y_1(0) - y_1(\varphi_0) + \frac{1}{gk\alpha_1\Delta_2 b_2}\arctan\left(\frac{1-2b_2\tan\varphi}{\Delta_2}\right)\bigg|_0^{\varphi} - \frac{1}{2gk\alpha_1 b_2}\ln\left(b_2\tan^2\varphi - \tan\varphi + b_1\right)\bigg|_0^{\varphi},$$

$$t_2(\varphi) = t_1(0) - t_1(\varphi_0) + \frac{1}{g\sqrt{-k\alpha_2}}\arcsin\left(\frac{2b_2\tan\varphi-1}{\Delta_4}\right)\bigg|_0^{\varphi}.$$

On the third interval we have

$$x_3(\varphi) = x_1(0) - x_1(\varphi_0) + x_2(\varphi_1) - x_2(0) - \frac{2}{gk\beta_1 d_2}\arctan\left(\frac{2d_1\tan\varphi - 1}{d_2}\right)\bigg|_{\varphi_1}^{\varphi} -$$

$$- \frac{w}{g\sqrt{k\beta_2}}\ln\left(1 - 2d_1\tan\varphi - 2\sqrt{d_1}\sqrt{d_0 - \tan\varphi + d_1\tan^2\varphi}\right)\bigg|_{\varphi_1}^{\varphi},$$

$$y_3(\varphi) = y_1(0) - y_1(\varphi_0) + y_2(\varphi_1) - y_2(0) - \frac{2}{gk\beta_2 d_2} \arctan\left(\frac{2d_1 \tan\varphi - 1}{d_2}\right)\bigg|_{\varphi_1}^{\varphi} -$$

$$-\frac{1}{2gk\beta_2} \ln\left(d_0 - \tan\varphi + d_1 \tan^2\varphi\right)\bigg|_{\varphi_1}^{\varphi}, \tag{10}$$

$$t_3(\varphi) = t_1(0) - t_1(\varphi_0) + t_2(\varphi_1) - t_2(0) - \frac{1}{g\sqrt{k\beta_2}} \ln\left(1 - 2d_1\tan\varphi - 2\sqrt{d_1}\sqrt{d_0 - \tan\varphi + d_1\tan^2\varphi}\right)\bigg|_{\varphi_1}^{\varphi}.$$

Index value $i$ in functions $x_i(\varphi), y_i(\varphi), t_i(\varphi)$ corresponds to the number of the movement interval $(i = 1, 2, 3)$. In relations (10), the following notation is introduced:

$$b_1 = \frac{1}{\alpha_1}\left(\frac{1}{ku_0^2 \cos^2\varphi_0} + f(\varphi_0)\right), \quad b_2 = \frac{\alpha_2}{\alpha_1}, \quad \Delta_1 = \sqrt{-1 - 4b_1b_2}, \quad \Delta_2 = \sqrt{-1 + 4b_1b_2},$$

$$\Delta_3 = \sqrt{1 + 4b_1b_2}, \quad \Delta_4 = \sqrt{1 - 4b_1b_2}, \quad \beta_2 = \frac{\left(f(\varphi_1) + f(89°)\right)\cos\varphi_1 - 2\left(\tan\varphi_1 + \tan 89°\right)}{(\tan\varphi_1 + \tan 89°)^2 \cos\varphi_1},$$

$$\beta_1 = \frac{2(1 + \beta_2 \sin\varphi_1)}{\cos\varphi_1}, \quad \beta_0 = f(\varphi_1) - \beta_1 \tan\varphi_1 + \beta_2 \tan^2\varphi_1, \quad d_0 = \frac{1}{\beta_1}\left(\frac{1}{ku_0^2\cos^2\varphi_0} + f(\varphi_0) - \beta_0\right),$$

$$d_1 = \frac{\beta_2}{\beta_1}, \quad d_2 = \sqrt{4d_0 d_1 - 1}.$$

Thus, on each of the intervals, the movement of the projectile is described by the equations:

at $i = 1$  $x = x_1(\varphi), \quad y = y_1(\varphi), \quad t = t_1(\varphi);$
at $i = 2$  $x = x_2(\varphi), \quad y = y_2(\varphi), \quad t = t_2(\varphi);$
at $i = 3$  $x = x_3(\varphi), \quad y = y_3(\varphi), \quad t = t_3(\varphi).$

Collectively, these equations describe the motion of the projectile over the entire interval of change in the trajectory angle $\varphi_0 \geq \varphi \geq -\pi/2$. If during the motion of the projectile the trajectory angle $\varphi$ is within the limits $\varphi_0 \geq \theta \geq \varphi_1$, then the functions $x_3(\varphi), y_3(\varphi), t_3(\varphi)$ are not used to describe of the movement. The projectile trajectory is given parametrically by the functions $x(\varphi), y(\varphi)$.

Formulas (10) are an improved version of the previously obtained formulas [4]. At very small values of the drag coefficient ($k = 10^{-12}$) and in the absence of wind, formulas (10) are transformed into formulas of the theory of parabolic projectile motion. The value $k = 0$ cannot be used in formulas (10), since division by zero occurs. Using formulas for $x_1(\varphi), y_1(\varphi), t_1(\varphi)$, we can find the values of the $x, y$ coordinates and time $t$ corresponding to the top of the projectile trajectory. Assuming $\varphi = 0$ in these formulas, we have

$$x_a = x_1(0) - x_1(\varphi_0), \quad H = y_1(0) - y_1(\varphi_0), \quad t_a = t_1(0) - t_1(\varphi_0).$$

Here $x_a$ is the abscissa of the vertex of the trajectory, $H$ is the maximum height of the projectile, $t_a$ is the projectile rise time (see Fig. 1).

### 4. Results of the calculations

To check the applicability of formulas (10), we consider the movement of various sports equipment – a golf ball, a tennis ball, a badminton shuttlecock. The drag coefficient $k$ for these projectiles is usually determined through the terminal velocity $V_{term}$ [5]:

$$k = \frac{1}{V_{term}^2} = const.$$

**Table 1. Air resistance coefficients and initial speed in calculations.**

| sport | $V_{term}$, m/s | $Re \times 10^5$ | $k$, s$^2$/m$^2$ | $V_0$, m/s |
|---|---|---|---|---|
| badminton | 6.7 | 0.27 | 0.022 | 60 |
| tennis | 22 | 0.95 | 0.002 | 50 |
| golf | 32.09 | 0.90 | 0.000971 | 40, 60 |

The values of the drag coefficient $k$ used in the calculations differ by a factor of 22. All of the figures below show the trajectories of the projectile. The thick solid black lines are obtained by numerical integration of system (5) with the aid of the 4-th order Runge-Kutta method (RK4). The red dot lines are obtained using analytical formulas (10).

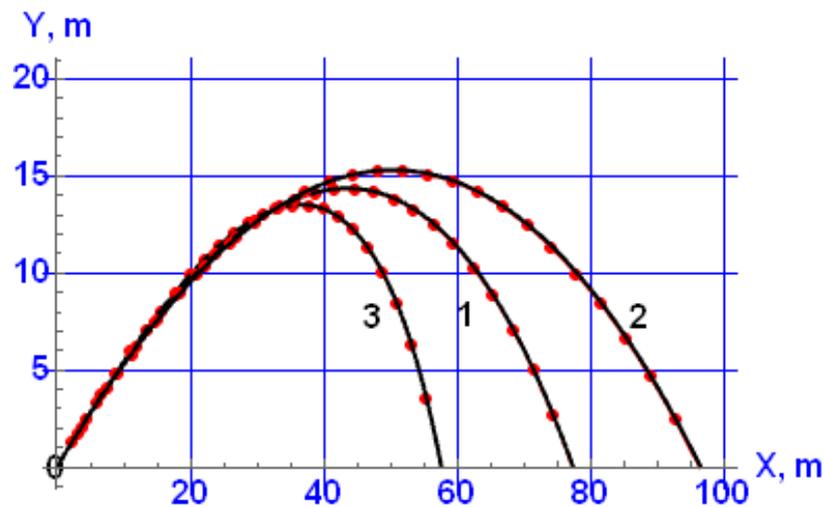

**Fig. 2.** Golf balls movement with parameters $V_0 = 40$ m/s, $g = 9.81$ m/s$^2$, $k = 0.000971$ s$^2$/m$^2$, $\theta_0 = 30°$.

Figure 2 reproduces Figure 4b from the paper [6]. Curve 1 corresponds to the movement of a golf ball without wind, curve 2 – to movement with a tailwind $w = 10$ m/s, curve 3 – the movement of the ball with a headwind $w = -10$ m/s. Figure 2 shows the effect of wind on the parameters $H$, $x_a$ and $L$ of the projectile.

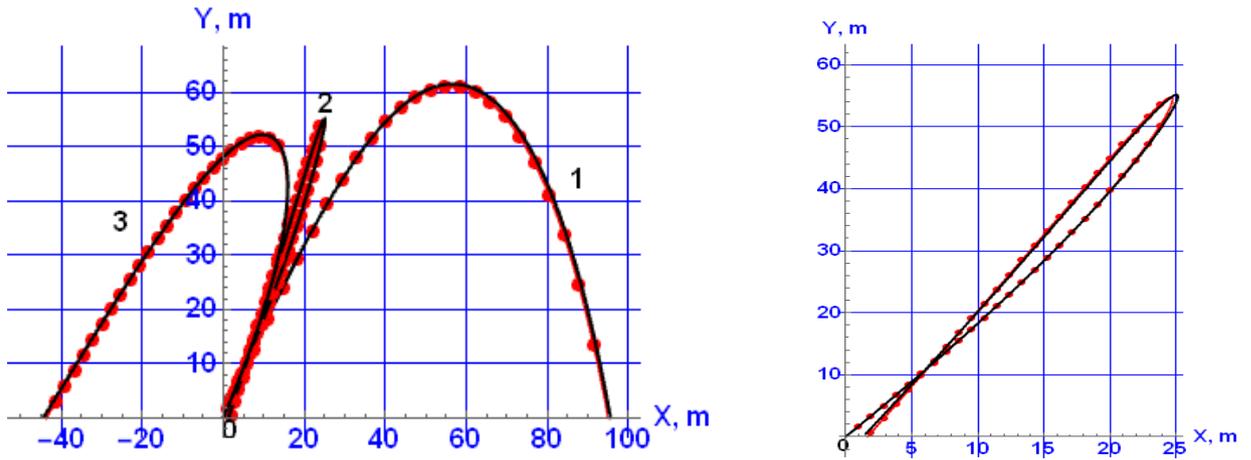

**Fig. 3.** Parameters for throwing a golf ball: $V_0 = 60$ m/s; $\theta_0 = 60°$ ; $k = 0.000971$ s$^2$/m$^2$. Trajectory 1 was plotted with $w = 0$ (no wind); trajectory 2 – with $w = -20$ m/s; trajectory 3 – with $w = -30$ m/s.

Figure 3 shows the influence of the headwind on the shape of the trajectory and the position of the end point of the trajectory – to the right or left of the throwing point. Trajectory 2 from the left picture of Figure 3 has a complex form with self-intersection and is shown in the right picture of Figure 3.

Of all the trajectories of sport projectiles, the trajectory of the shuttlecock has the greatest asymmetry. This is explained by the relatively large value of the drag coefficient $k$. In addition, the trajectory of the shuttle approaches to the vertical asymptote very quickly.

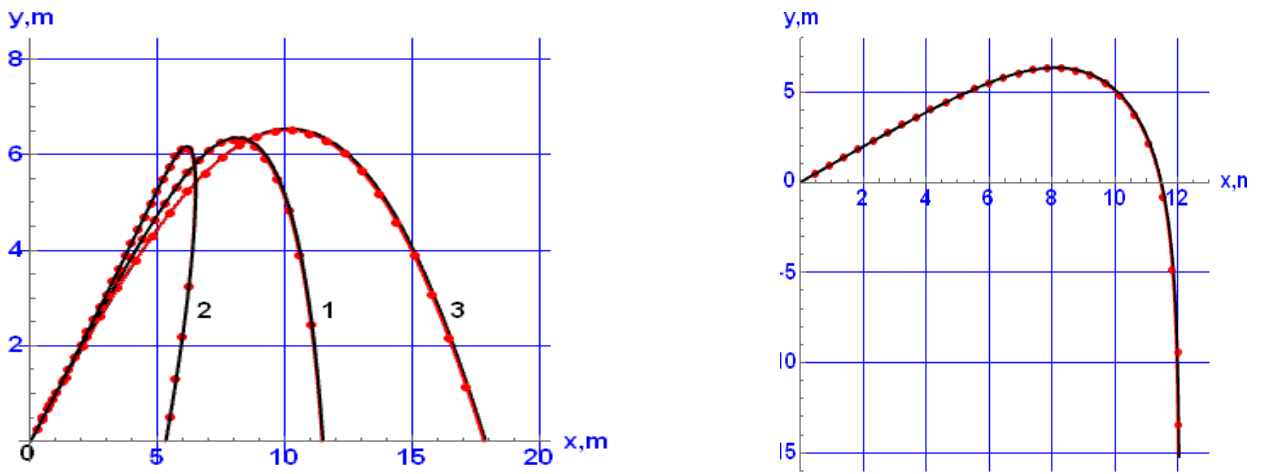

**Fig. 4.** Badminton shuttlecock throwing parameters: $V_0 = 60$ m/s; $\theta_0 = 45°$, $k = 0.022$ s$^2$/m$^2$. On the left picture curve 1 was plotted at the value $w = 0$ (not wind), curve 3 – at $w = 3$ m/s, ; curve 2 – at $w = -3$ m/s.

In the right picture of Figure 4 shows the trajectory of the shuttlecock when moving along the asymptote without wind. Figure 4 shows that the trajectory of the shuttlecock constructed using formulas (10) practically coincides with the vertical asymptote. The author is not aware of any approximate analytical solution of the projectile motion problem that would describe the motion along the asymptote. Figure 5 shows the oblique asymptote for a wind speed of 10 m/s.

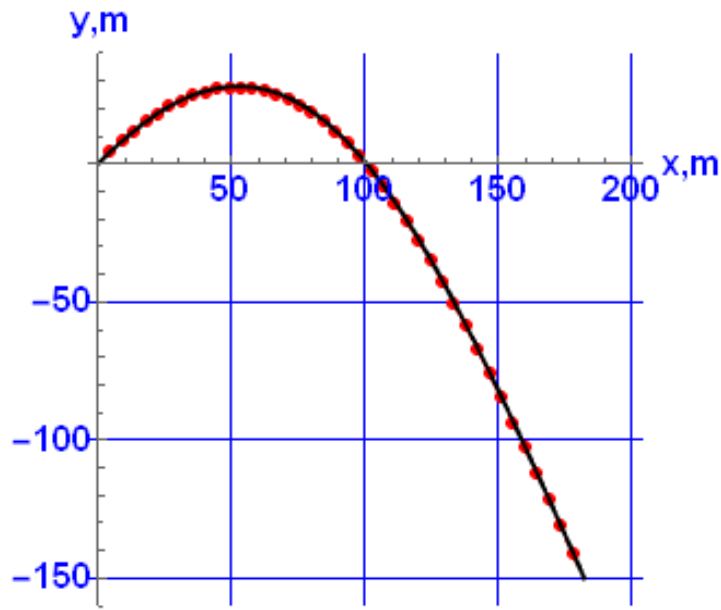

**Fig. 5.** Tennis ball throwing parameters: $V_0 = 50$ m/s; $\theta_0 = 45°$, $k = 0.002$ s$^2$/m$^2$, $w = 10$ m/s.

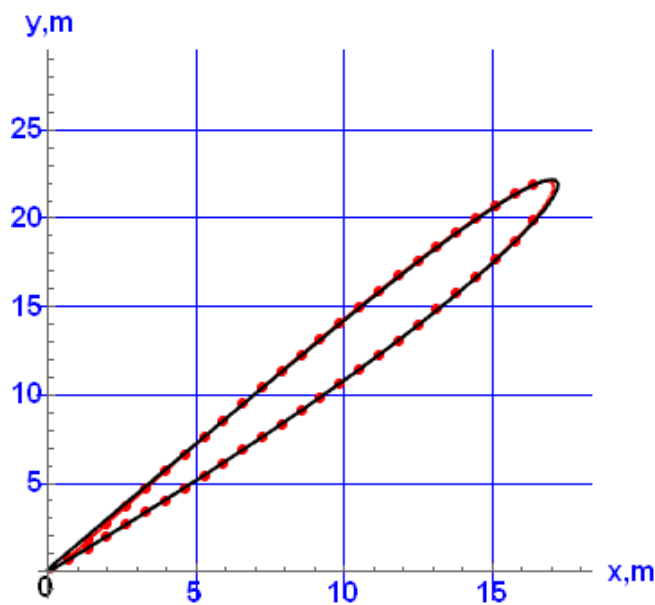

**Fig. 6.** Tennis ball throwing parameters: $V_0 = 50$ m/s; $\theta_0 = 45°$, $k = 0.002$ s$^2$/m$^2$, $w = -19.75$ m/s.

Figure 6 shows the trajectory of a tennis ball at the corresponding value of the wind speed. The ball returns to the point of throw at the selected value of the wind speed. All presented figures demonstrate excellent agreement between numerical solutions RK4 (solid black curves) and analytical solutions (red dotted curves), which are described by the formulas (10).

## 5. Conclusions

As examples of the use of formulas (10), the movement of various objects was considered – a golf ball, shuttle of badminton, tennis ball. The calculation results testify to the universality of the proposed analytical solutions (10), which are an improved version of formulas [4]. They are operable over a wide range of initial throw conditions and drag coefficients. The relative maximum deviation of the analytical value (10) from the numerical value (RK4) at any point of the trajectory does not exceed 1%. At very small values of the drag coefficient, formulas (10) are transformed into formulas of the theory of parabolic projectile motion. It should be noted the educational and methodological benefits of solutions (10) for introductory physics courses studied by students of colleges and universities